\documentclass[a4paper]{jpconf}
\usepackage{graphicx}
\usepackage{iopams}

\usepackage{color}

\def\be{\begin{equation}}
\def\beq{\begin{equation}}
\def\eeq{\end{equation}}
\def\ee{\end{equation}}
\def\bea{\begin{eqnarray}}
\def\eea{\end{eqnarray}}
\def\ba{\begin{array}}
\def\ea{\end{array}}

\def \ga {\gamma}

\def\IZ{\mathbb{Z}}
\def\IR{\mathbb{R}}

\def\IT{\mathbb{T}}

\def\rref#1{(\ref{#1})}

\begin{document}
\title{A quantum Goldman bracket in $2+1$ quantum gravity}\footnote{Dedicated to John Charap, in gratitude for his inspiration and vision.}

\author{J E Nelson ${}^1$ and R F Picken ${}^2$}

\address{${}^1$ Dipartimento di Fisica Teorica, Universit\`a
degli Studi di Torino and Istituto Nazionale di Fisica Nucleare, Sezione di
Torino,  via Pietro Giuria 1, 10125 Torino, Italy}
\address{ ${}^2$ Departamento de Matem\'{a}tica and CAMGSD - Centro de An\'{a}lise Matem\'{a}tica, Geometria e Sistemas Din\^{a}micos,
 Instituto Superior T\'{e}cnico, Technical University of Lisbon,
Avenida Rovisco Pais, 1049-001 Lisboa, Portugal}

\ead{nelson@to.infn.it, rpicken@math.ist.utl.pt}

\begin{abstract}
In the context of quantum gravity for spacetimes of dimension  $2+1$, we describe progress in the construction of a quantum Goldman bracket for intersecting loops on surfaces. Using piecewise linear paths in $\mathbb{R}^2$ (representing loops on the spatial manifold, i.e. the torus) and a quantum connection with noncommuting components, we review how holonomies and Wilson loops for two homotopic paths are related by phases in terms of the signed area between them. Paths rerouted at intersection points with other paths occur on the r.h.s. of the Goldman bracket. To better understand their nature we introduce the concept of integer points inside the parallelogram spanned by two intersecting paths, and show that the rerouted paths  must necessarily pass through these integer points. 
\end{abstract}

\section{Introduction}

\noindent In previous work \cite{NP1,mod,qmp} we have investigated quantum gravity in $2+1$ dimensions with negative cosmological constant on the torus, using an approach involving quantum holonomy matrices. This followed on from earlier work by one of us with Regge and Zertuche \cite{NR1,NRZ} based on the traces of the holonomies. In \cite{goldman} we focused on the quantum geometry that arises from introducing a constant quantum connection, from which the holonomy matrices and their traces are obtained (in the sector where these matrices are diagonal). Some interesting features emerged, in particular a quantum version of the well-known Goldman bracket for loops on a surface. In the present article we describe some new developments in  the understanding of this quantum geometrical picture.

The classical action of $2+1$ gravity with negative cosmological constant $\Lambda$, in the dreibein formulation, was related by Witten \cite{wit} to Chern-Simons theory for the gauge group $SL(2,\IR)\times SL(2,\IR)$. Thus the classical phase space of this model of $2+1$ gravity corresponds to the moduli space of flat $SL(2,\IR)\times SL(2,\IR)$ connections on the spatial manifold, which we take to be the torus $\IT^2=\IR^2/\IZ^2$. Treating each $SL(2,\IR)$ factor separately, we are then led to consider pairs of commuting $SL(2,\IR)$ matrices (up to simultaneous conjugation by an element of $SL(2,\IR)$), representing the holonomies of the flat connection along two generating cycles of the torus (up to gauge transformation). Note that the fundamental group of the torus is generated by two cycles, with a single relation saying that the generators commute. For a detailed description of the moduli space of flat $SL(2,\IR)$ connections on the torus, see \cite{mod}.

The diagonal sector of this moduli space
can be parametrised by constant connections \cite{mikpic,goldman} on the torus of the form
\be
A= (r_1 dx + r_2 dy)
\left(\begin{array}{clcr} 1& 0 \\0& -1 
\end{array}\right)   
\label{conn}
\ee
where $x,\,y$ are coordinates on the torus (with periodicity $1$), or on its covering space $\IR^2$. The connections are constant in the sense that the functions $r_1,\,r_2$ are constant functions (independent of $x$ and $y$). After integration, this yields the holonomies:
\be
U_i = \exp \int_{\ga_i} A = \left(\begin{array}{clcr}e^{{r_i}}& 0 \\0& e^{-
{r_i}}\end{array}\right) \quad i=1,2
\label{hol1}
\ee
where $\gamma_1,\, \gamma_2$ are generating cycles of the fundamental
group of the torus.

In Chern-Simons theory components of the connection $A$ have non-trivial Poisson brackets
amongst themselves. At the level of the connection components $r_i,\,i=1,2$ this
translates into the Poisson bracket
\be
\left\{ r_1, r_2\right\} = -\frac{\sqrt {-\Lambda}}{4},
\ee
which is quantised by 
\begin{equation}
[\hat{r}_1, \hat{r}_2] = \frac{i\hbar \sqrt{-\Lambda}}{4}.
\label{rcommrel}
\end{equation}
Therefore we will consider a quantised phase space parametrised by constant quantum connections of the form:
\be
\hat{A}= (\hat{r}_1 dx + \hat{r}_2 dy) 
\left(\begin{array}{ll} 1& 0 \\0& -1 \end{array}\right).
\label{quconn}
\ee

After integration along the generating cycles the Wilson loops, i.e. traces of the holonomies, of this connection are precisely the variables used by Nelson and Regge \cite{NR1}. This connection also yields the description in terms of quantum holonomy matrices of Nelson and Picken \cite{NP1,qmp}. We remark also that the relation between the parameters $r_1,\,r_2$ (a pair for each $SL(2,\IR)$ factor) and the ADM variables for $2+1$ gravity has been elucidated in \cite{CN}. Our purpose here is to further explore the quantum geometry arising from the assignment of a quantum matrix (or its trace) to a general class of loops on the torus.

\section{Quantum geometry of quantum holonomy matrices}

It is convenient to identify loops on the torus $\mathbb{T}^2$ with paths on its covering space $\mathbb{R}^2$. We therefore consider all loops on the torus as being represented by piecewise-linear (PL) paths between integer points on the $(x,y)$ plane, and work with this description, bearing in mind that paths upstairs i.e. in $\mathbb{R}^2$ represent loops downstairs i.e. on the torus $\mathbb{T}^2$. In particular, a natural class of paths $p$ to consider are those straight paths that start at the origin and end at an integer point $(m,n)\in \IZ^2$. They will be denoted $p=(m,n)$.
Using the quantum connection \rref{quconn} we assign a quantum matrix to any such straight path by
\be
\hat{U}_{(m,n)}= \exp \int_{(m,n)} \hat{A} = 
\left(\begin{array}{ll} e^{m\hat{r}_1+ n\hat{r}_2} & 0 
\\ 0 & e^{-m\hat{r}_1 -n\hat{r}_2}
\end{array}\right).
\label{Umn}
\ee
This assignment is straightforwardly extended to any PL path between integer points by assigning a quantum matrix to each linear segment of the path, in an analogous fashion to \rref{Umn}, and multiplying the matrices in the same order as the segments appear along the path. This prescription obviously coincides with the general relation:
\be
p\mapsto \hat{U}_p = {\cal P }\exp \int_{p} \hat{A}.
\label{hol}
\ee
where $\cal P$ denotes path-ordering.

In the upstairs picture, two homotopic loops on the torus are represented by two PL paths on the plane, $p_1,\, p_2$, with the same integer starting point and the same integer endpoint.  It was shown in \cite{goldman} that the following relationship holds for
the respective quantum matrices:
\begin{equation}
\hat{U}_{p_1}=q^{S(p_1,p_2)}\hat{U}_{p_2},
\label{areaphase}
\end{equation}
where $S(p_1,p_2)$ denotes the signed area enclosed between the paths
$p_1$ and $p_2$, and 
$
q=e^{-i \hbar \sqrt{-\Lambda}/4}.
$

\begin{figure}
\begin{center}
\includegraphics[width=16pc]{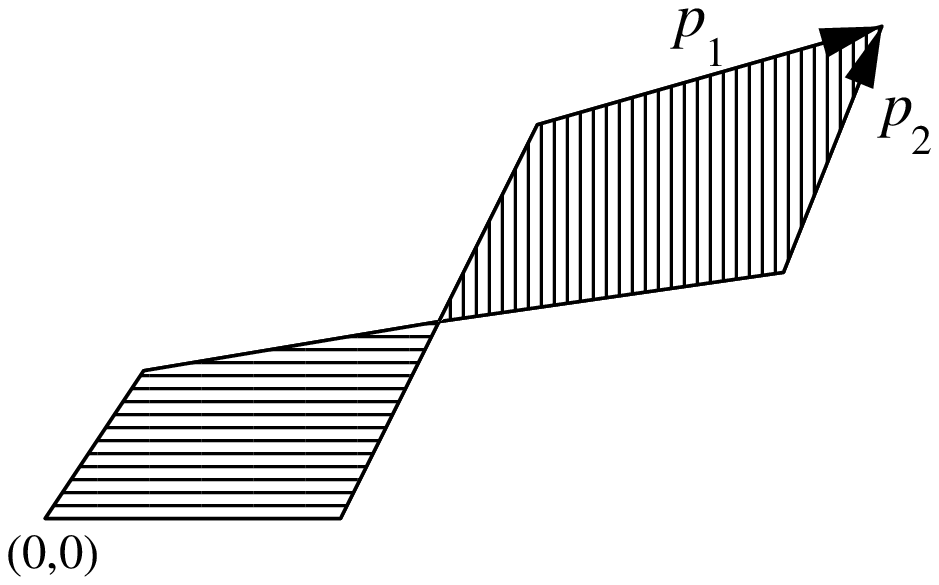}
\end{center}
\caption{\label{signedarea} The signed area between $p_1$ and $p_2$ is the area of the the horizontally shaded region minus the area of the vertically shaded region}
\end{figure}

The signed area between two such PL paths is defined as follows: for any finite region $R$ enclosed by $p_1$ and $p_2$, if the boundary of $R$ consists of oriented segments of $p_1$ and $p_2^{-1}$ (the orientation reversal of $p_2$), and as such is globally oriented in the positive (anticlockwise), or negative (clockwise) sense, this gives a contribution of $+{\rm area}(R)$ , or $-{\rm area}(R)$ respectively, to the signed sum $S(p_1,p_2)$ (otherwise the contribution is zero). See Figure \ref{signedarea}.

We note for the discussion in the next sections that the Wilson loops, or traces of the holonomies, are related by the same ``area phases'', i.e.
\begin{equation}
\hat{T}({p_1})=q^{S(p_1,p_2)}\hat{T}({p_2}),
\label{areaphasewilson}
\end{equation}
where $\hat{T}(p)= {\rm tr}~\hat{U}_{p}$.

In \cite{goldman} this relation between the holonomy matrices for homotopic loops/paths was interpreted as a $q$-deformed surface group representation, i.e. a deformation of a representation of the fundamental group in terms of matrices. There we used the fact that signed area between paths has the following two properties:
\begin{equation}
 S(p_1,p_3) =  S(p_1,p_2) + S(p_2,p_3), \quad\quad S(p_1p_2,p_3p_4) = S(p_1,p_3) + S(p_2,p_4)
\label{signedareaprops}
\end{equation}
where $p_1p_2$ denotes the concatenation of paths $p_1$ and $p_2$.
Another perspective on this comes from recent work by one of us with Martins \cite{FMP}, where the notion of $2$-dimensional holonomy was explored. In the present context, $2$-dimensional holonomy may be described as the assignment of an element of a group $G$ to each of a pair of homotopic loops, and simultaneously the assignment of an element of a group $E$ to the homotopy between the two loops, subject to certain consistency relations. Amongst these relations, there is the requirement that the elements assigned to the homotopies behave well under vertical and horizontal composition, which translates precisely to the properties of signed area as in \rref{signedareaprops}. We note also that a crucial ingredient in the construction of \cite{FMP} is a connection $2$-form with values in the Lie algebra of $E$, and indeed in the present context there is a natural $2$-form, namely the non-vanishing curvature of the connection  \rref{quconn} (non-vanishing because of the non-commutativity of the components $r_1$ and $r_2$, as pointed out in \cite{goldman}). Thus there are strong hints that the quantum holonomies of $2+1$ gravity may be interpreted in terms of $2$-dimensional holonomy.

\section{Quantum geometry of Wilson loops and the quantum Goldman bracket}
When analysing the behaviour of the Wilson loops $\hat{T}(p)= {\rm tr}~\hat{U}_{p}$ for the quantum connection \rref{quconn} in \cite{goldman} a link with the Goldman bracket \cite{gol} emerged. This bracket is a Poisson bracket for functions  $T(\gamma)= {\rm tr}\, U_\ga$, defined on homotopy classes of loops on a surface, which for $U_\ga \in SL(2,\IR)$ takes the following form (see
\cite{gol} Thm. 3.14, 3.15 and Remark (2), p. 284):
\be
\{T(\ga_1), T(\ga_2)\} = \sum_{S \in \ga_1 \sharp \ga_2}
\epsilon(\ga_1,\ga_2,S)(T(\ga_1S\ga_2) -
T(\ga_1S\ga_2^{-1})).
\label{gold}
\ee
Here $\ga_1 \sharp \ga_2$ denotes the set of (transversal) intersection points
of $\ga_1$ and $\ga_2$ and $\epsilon(\ga_1,\ga_2,S)$ is the intersection number
of $\ga_1$ and $\ga_2$ at the intersection point $S$, i.e. (for simple intersections)
 $+1$ if the angle between the tangent
vector of $\ga_1$ at $S$ and the tangent
vector of $\ga_2$ at $S$ is between $0$ and $180$ degrees, and $-1$ if it is between
$180$ and $360$ degrees.
Finally
 $\ga_1S\ga_2$ and $\ga_1S\ga_2^{-1}$ denote
loops which are  {\it rerouted} at the intersection point $S$, meaning that we follow $\ga_1$ as far as the intersection point $S$, then follow $\ga_2$ (or the inverse loop $\ga_2^{-1}$) from $S$ back to $S$, and finally proceed along $\ga_1$ from $S$ back to the starting point of $\ga_1$. For a fuller discussion and examples see section 5 of \cite{goldman}. 

We found that the Wilson loops $\hat{T}(p)$, for $p=(m,n)$ a straight path corresponding to a loop on the torus, satisfied a quantum version of the Goldman relation \rref{gold}. In fact two different quantisations emerged, a straightforward one, and a refined one, which we will now describe.

First, for straight paths $p_1=(m,n)$ and $p_2=(s,t)$, the Goldman bracket takes the form:
\be 
\left\{T(m,n), T(s,t)\right\} = (mt-ns)
(T(m+s,n+t)- T(m-s,n-t)). 
\label{straightforward}
\ee 
Here $mt-ns$ is the total intersection number between $p_1$ and $p_2$, and this factor appears because there are effectively $mt-ns$ simple intersection points and the corresponding reroutings  $p_1Sp_2$ are all homotopic to the straight path $(m+s, n+t)$, with an analogous statement for the negative reroutings $p_1Sp_2^{-1}$. 

By a simple direct calculation, it was shown in \cite{goldman} that the Wilson loops satisfy:
\begin{equation}
[\hat{T}(m,n), \hat{T}(s,t)]=
(q^{\frac{mt-ns}{2}}-q^{-\frac{mt-ns}{2}}) \left(\hat{T}(m+s,n+t) - \hat{T}(m-s,n-t)\right)
\label{qgb1}
\end{equation}
i.e. a quantisation of \rref{straightforward}, with the total intersection number $mt-ns$ replaced by a quantum total intersection number (the first factor on the r.h.s. of \rref{qgb1}). We will call this the straightfward quantisation of the Goldman bracket.

A more refined quantisation was also obtained in \cite{goldman}, where each rerouting appears as a separate term, and the area phases of these different but homotopic reroutings are taken into acccount. This takes the following form:
\be
[\hat{T}(p_1), \hat{T}(p_2)] = \sum_{ {S} \in p_1 \sharp p_2}
(q^{\epsilon(p_1,p_2,{S})} - 1)\hat{T}(p_1{S}p_2)  
+ (q^{-\epsilon(p_1,p_2,{S})} - 1)\hat{T}(p_1{S}p_2^{-1})
\label{qgold}
\ee
which quantises \rref{gold} (with loops $\ga$ substituted by paths $p$) by replacing the intersection numbers $\epsilon(p_1,p_2,{S})$ by quantum intersection numbers $(q^{\epsilon(p_1,p_2,{S})} - 1)$. 

\begin{figure}[hbpt]
\centering
\includegraphics[height=4cm]{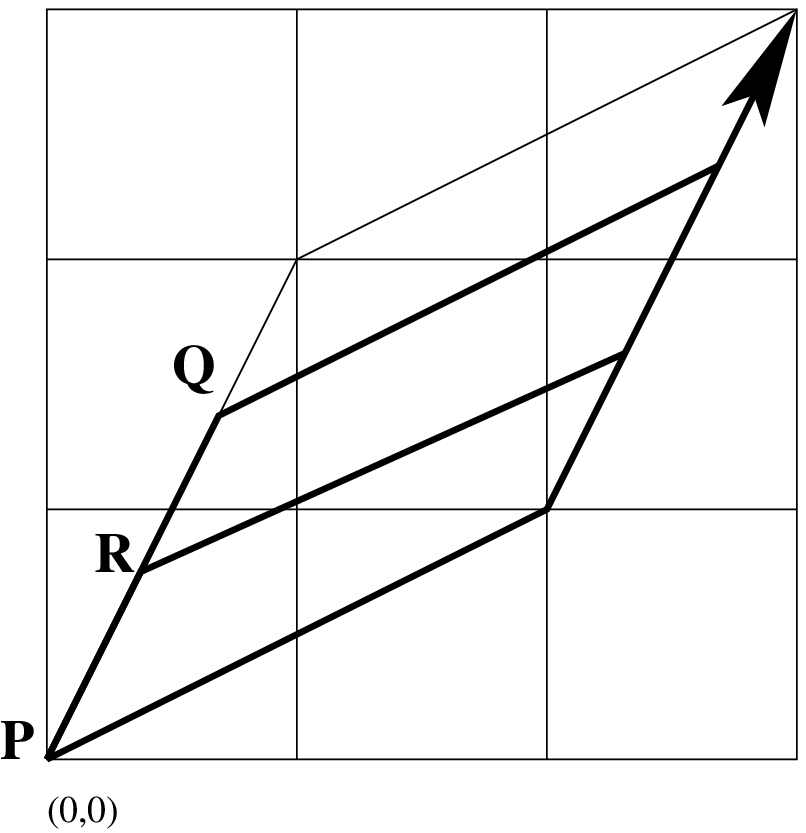}
\hspace{2cm}
\includegraphics[height=4cm]{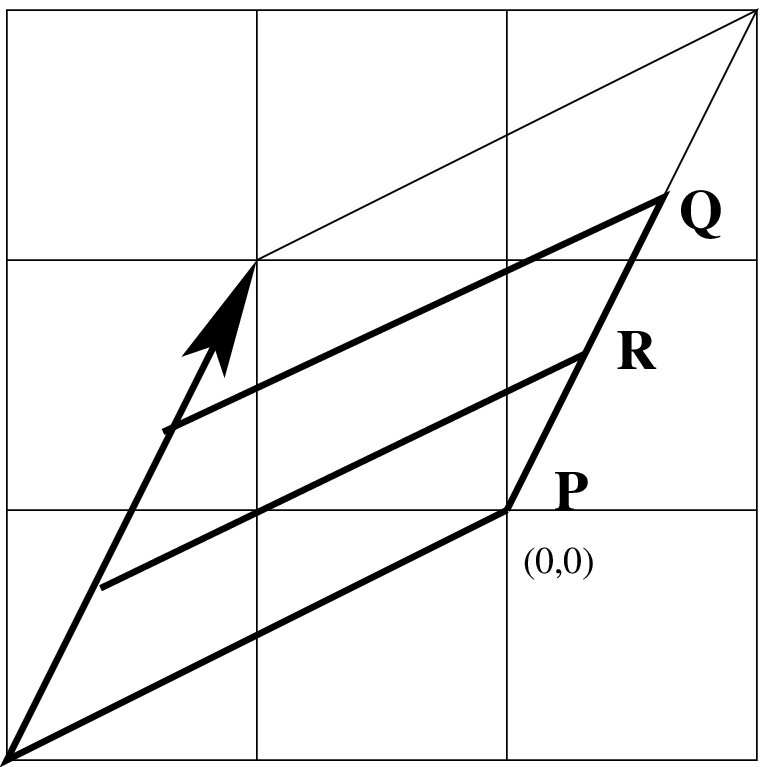}
\caption{ The reroutings $(1,2)S(2,1)$ and $(1,2)S(-2,-1)$ for $S=P,R,Q$ }
\label{p20a}
\end{figure}

An example is given by the choice $p_1=(1,2)$, $p_2=(2,1)$, where the refined bracket is given by
\be
[\hat{T}(1,2),\hat{T}(2,1)] =\sum_{S=P,R,Q} 
(q^{-1} -1) \hat{T}((1,2)S(2,1)) + (q-1) \hat{T}((1,2)S(-2,-1)), 
\label{qgoldexp}
\ee
where the terms on the r.h.s. come from the positive and negative reroutings at three intersection points denoted $P,\,R,\,Q$, see Figure \ref{p20a}. We then calculate the area phases of each rerouting relative to the corresponding straight path $(3,3)$ or $(-1,1)$, and using \rref{areaphasewilson} we obtain:
\begin{eqnarray}
[\hat{T}(1,2),\hat{T}(2,1)] &=& (q^{-1}-1)(q^{\frac{3}{2}} + q^{\frac{1}{2}} + q^{-\frac{1}{2}}) \hat{T}(3,3)
+ (q-1) (q^{-\frac{3}{2}} + q^{-\frac{1}{2}} + q^{\frac{1}{2}}) \hat{T}(-1,1) \nonumber \\
&=& (q^{-\frac{3}{2}} - q^{\frac{3}{2}}) (\hat{T}(3,3) - \hat{T}(-1,1)).
\end{eqnarray}
i.e. the straightforward form of this commutator.

In the next section we describe some new insights which allow a better understanding of intersections and reroutings, and prepare the way to address some outstanding issues concerning the two quantisations \rref{qgb1} and \rref{qgold} (see the final section).

\section{Intersection points and integer points}
In \cite{goldman} we observed that the paths that appear after rerouting two straight paths at an intersection point all pass through an integer point in the parallelogram spanned by the two paths. Here we will show from a new viewpoint that this is necessarily the case, and we shall see that it is also a very useful way to examine the nature of the reroutings.

We look at reroutings of the type $p_1Rp_2$, where $p_1$ and $p_2$ are both straight paths in $\IR^2$ starting at the origin, corresponding to loops on the torus $\IT^2= \IR^2/\IZ^2$. This means that we follow
$p_1$ up to the intersection point $R$, reroute along $p_2$ starting at the point $R$
until $p_2$ returns to $R$, and complete the remainder of the first 
 path $p_1$ from $R$ back to the origin. Note that the intersection point $R$ may occur at the origin of $\IR^2$;  in that case we follow $p_2$ straight away. Instead of studying the
 intersections between $p_1$ and $p_2$ by reducing both paths to a
 fundamental domain, as we did in \cite{goldman}, we work directly in 
 $\IR^2$. Fix the path $p_1$ and 
 parallel translate $p_2$ to start at a new integer point, denoted $\alpha$, in such a
 way that it intersects $p_1$ at $R$. Of course, all integer points are the
 same when projecting  down to the torus, but this shift of $p_2$ gives a very
 clear picture of where $R$  is located along both
 paths simultaneously. 

We introduce the following notation for paths on the plane: $(p)_{A,B}$ denotes the subpath of the path $p$ going from the point $A$ on $p$ to the point $B$ on $p$. Also for any path $p$, let $\overline{p}$ denote the integer endpoint of $p$. Note that for straight paths $p=(m,n)$, the notation for the endpoint and for the path itself coincide, i.e. $\overline{p}=p=(m,n)$,  but when the path is non-straight, there is a distinction between $p$ and $\overline{p}$. This latter situation will occur when we try and extend the refined bracket to non-straight paths - see the outstanding issues listed in the final section.

\begin{figure}
\begin{center}
\includegraphics[width=12pc]{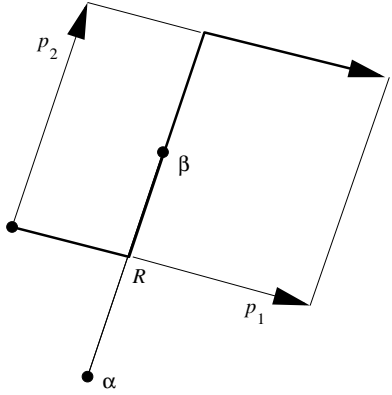}
\end{center}
\caption{\label{rerouting} The rerouting $p_1Rp_2$ }
\end{figure}

Now write $p_1= (p_1)_{0R}(p_1)_{R\overline{p_1}}$, to indicate how $p_1$ is divided into two segments by the intersection point $R$. Here $p_1$ starts at the origin $0$, but the parallel translated $p_2$ starts at the integer point $\alpha$, so we require a notation to describe the shift or parallel translation of paths to start at different points: $p^\alpha$ will denote the path $p$ parallel translated to start at $\alpha$ instead of its original starting point. (Of course, projected down to the torus, the original $p$ going from the origin to an integer point, and the shifted path $p^\alpha$ give rise to identical loops). In this way, we can write down an explicit algebraic expression for the rerouting $p_1Rp_2$ represented as a path in the plane (see Figure \ref{rerouting}):
\[
p_1Rp_2 = (p_1)_{0,R}\, (p_2^\alpha)_{R,\beta}\, (p_2^\beta)_{\beta, R+\overline{p_2}} \,
(p_1^{\overline{p_2}})_{R+ \overline{p_2},\overline{p_1} +\overline{p_2}}, 
\]
where the start and endpoints $\alpha$ and $\beta$ of $p_2^\alpha$ are related by: $\beta = 
\alpha + \overline{p_2}$. 

This detailed notation for rerouted paths should prove useful in the future for manipulating paths algebraically and making precise general statements. However the approach in this section also makes it very clear why intersections occur and how they give rise to reroutings. In concrete terms, for an intersection to occur, we need either a starting point $\alpha$ for $p_2^\alpha$ such that $p_2^\alpha$ intersects $p_1$ in a point $R$ (which may be the origin, but not the endpoint of $p_1$), or equivalently we need an endpoint $\beta$ such that the appropriate $p_2^\alpha$ ending in $\beta$ intersects $p_1$ in a point $R$ as before.

\begin{figure}
\begin{center}
\includegraphics[width=12pc]{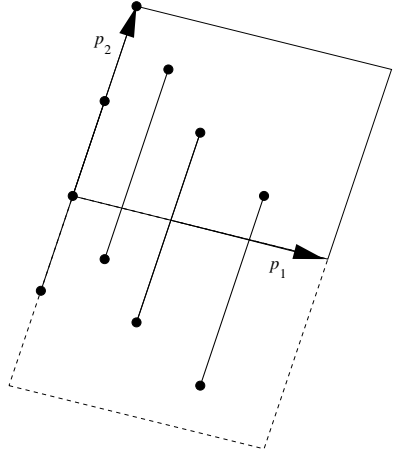}
\end{center}
\caption{\label{preparallelogram} The parallelogram and pre-parallelogram (dotted line) for $p_1$ and $p_2$, showing some of the integer points (black dots) and the corresponding parallel-translated copies of $p_2$}
\end{figure}

Looking first at the possible starting points $\alpha$, we see that they are the integer points lying in a ``pre-parallelogram'' with vertices $-\overline{p_2}$, $-\overline{p_2} +\overline{p_1}$, $\overline{p_1}$ and the origin $0$. See Figure \ref{preparallelogram}. Here we should exclude any integer points lying on the edge between  
$-\overline{p_2} +\overline{p_1}$ and $\overline{p_1}$, since the corresponding paths $p_2^\alpha$ intersect $p_1$ at its endpoint, and we should also exclude any integer points lying on the edge between  $-\overline{p_2}$ and
$-\overline{p_2} +\overline{p_1}$, to avoid double counting, as we are including the integer points lying along the edge between $0$ and $\overline{p_1}$.

Equivalently, we can look at the endpoints $\beta$ lying inside the parallelogram generated by $p_1$ and $p_2$, i.e. with vertices $0$, $\overline{p_1}$, $\overline{p_1} + \overline{p_2}$ and $\overline{p_2}$. Here we will exclude the integer points lying on the edge between  $\overline{p_1}$ and
$\overline{p_1} +\overline{p_2}$ and those lying along the edge between $0$ and $\overline{p_1}$ (these endpoints correspond to the starting points we excluded from the pre-parallelogram).

From both perspectives we can readily verify that the number of intersections is correct, since the total intersection number is: 
\[ 
\epsilon (p_1,p_2) = \det \left(p_1 p_2\right).
\]
The total number of intersection points (counting multiplicities), all of which have the same sign of the intersection number, is therefore the modulus of $\epsilon (p_1,p_2)$, i.e. in geometric terms, the area of the parallelogram or the pre-parallelogram. In turn, this area is given by a classical theorem of Pick \cite{pick}, which states that the area $A(P)$ of a planar polygon $P$ with
vertices at integer points of the plane is given in terms of the number of
interior integer points $I(P)$ and the number of boundary integer points $B(P)$
as follows:
\[
A(P) = I(P) + \frac{B(P)}{2} -1.
\]
This value is evidently the same as the number of integer starting points $\alpha$ in the pre-parallelogram, or equivalently the number of integer end points $\beta$ in the parallelogram generated by $p_1$ and $p_2$, since we have excluded two edges, i.e. half the integer points lying along the interior of the edges and three out of four of the vertices (this is accounted for by the $-1$ in the formula, since $\frac{4}{2} -1 =1$, the single remaining vertex). 

A convenient feature of this approach is that it handles reducible paths in a natural manner. A reducible path is one such that $p=(m,n)=c(m',n')$, where $m,n,c,m',n'$ are all integers, with $c\geq 2$. If $p_2$ is reducible, this means that there are integer points along $p_2$ other than the endpoints. In terms of the pre-parallelogram analysis, there will be $c$ integer starting points $\alpha$ for each intersection point $R$ along $p_1$. The intersection point $R$ therefore has intersection number $\pm c$ and the rerouting at $R$ will appear with multiplicity $c$. These issues were briefly touched upon at the end of the penultimate section of \cite{goldman}.

A final point that emerges from this discussion is that, as we observed in \cite{goldman}, obviously any rerouting must pass through one of the integer points inside the parallelogram generated by $p_1$ and $p_2$ (namely the appropriate endpoint $\beta$). 

\section{Conclusions}
We have presented a new perspective on intersections and reroutings which holds promise of further progress on the following outstanding and interrelated issues: 

\begin{itemize}
 \item The quantum intersection numbers which naturally appear in the two quantisations  \rref{qgb1} and \rref{qgold} have a different appearance, and in particular they are symmetric, for the straightforward bracket, under the interchange $q \leftrightarrow q^{-1}$. This is not the case for the refined bracket.
\item Although there are strong arguments suggesting that, for straight paths, the refined bracket and the straightforward bracket are equivalent, a full proof is still lacking. 
\item Part of the previous problem is dealing with reducible paths, i.e. paths such that $p=(m,n)=c(m',n')$, where $m,n,c,m',n'$ are all integers, with $c\geq 2$.
\item We would like to extend the refined bracket beyond straight paths on the left to a more general class of paths, and ultimately get a definition of the refined bracket which closes on the class of paths.
\item The antisymmetry of the refined bracket for two straight paths is not manifest.
\item We must prove the Jacobi identity for the refined bracket extended to a wider class of paths.
\end{itemize}

In particular, the tools are now ready to start analysing the refined bracket \rref{qgold} when the paths on the l.h.s. are no longer straight. In the first instance we hope to address a phenomenon that occurred when we performed direct calculations to see how the refined bracket might extend to non-straight paths: we encountered some unexpected extra phases when the second path $p_2$ in the rerouting $p_1Rp_2$ was non-straight, and the first path $p_1$ was straight. These difficulties did not occur when the first path was non-straight and the second straight, i.e. the discrepancy was clearly related to the fact that the middle part of the rerouting was non-straight.

In future work we also hope to explore in more detail the link with $2$-dimensional holonomy \cite{FMP}, as discussed at the end of section 2, and gain further understanding of the elegant quantum geometry that emerges through the quantised Goldman bracket, relating it e.g. to noncommutative geometry \cite{con} or to other quantisations of the Goldman bracket \cite{Turaev}.
%loop quantum gravity programme - see \cite{thie} for a recent review

\medskip

\ack
This work was supported by the Istituto Nazionale di Fisica Nucleare
(INFN) of Italy, Iniziativa Specifica FI41, the Italian Ministero
dell'Universit\`a e della Ricerca Scientifica e Tecnologica (MIUR), and the {\em Programa Operacional Ci\^{e}ncia e Inova\c{c}\~{a}o 2010}, project number 
POCI/MAT/60352/2004, financed by the {\em Funda\c{c}\~{a}o para a Ci\^{e}ncia e a Tecnologia} (FCT) and cofinanced by the European Community fund FEDER.

\medskip

%\subsection{Appendices}

%%%%%%%%%%%%%%%%%%%%%%%%%%%%%%%%%%%%%%%%%%%


\section*{References}
\begin{thebibliography}{10}

\bibitem{NP1} Nelson J E and Picken R F 2000 Quantum holonomies in $(2+1)$-dimensional gravity {\it Phys. Lett.} B {\bf 471} 367--72.

\bibitem{mod} Nelson J E and Picken R F 2002 Parametrization of the
moduli space of flat  $SL(2,\IR)$ connections on the torus {\it Lett. Math.Phys.} {\bf 59} 215--26.

\bibitem{qmp} Nelson J E and Picken R F 2000 Quantum matrix pairs
{\it Lett. Math. Phys.} {\bf 52} 277--90.

\bibitem{NR1} Nelson J E and Regge T 1991 $2+1$ Quantum Gravity {\it Phys. Lett.} B {\bf
272} 213--16.

\bibitem{NRZ} Nelson J E, Regge T and Zertuche F 1990 Homotopy Groups and
$2+1$ dimensional Quantum De Sitter Gravity {\it Nucl. Phys.} B {\bf 339} 516--32.

\bibitem{goldman} Nelson J E and Picken R F 2005 Constant connections, quantum holonomies and the Goldman bracket, {\it Adv. Theor. Math. Phys.} {\bf 9} 3 407--33.

\bibitem{wit} Witten E 1988 2+1 dimensional gravity as an exactly soluble
system {\it Nucl. Phys.} B {\bf 311} (1988) 46--78.

\bibitem{mikpic} Mikovic A and Picken R F 2001 Super Chern-Simons theory and
flat super connections on a torus {\it Adv. Theor. Math. Phys.} {\bf 5} 243--63.

\bibitem{CN} Carlip S and Nelson J E 1995 Comparative quantisations of 2+1 Gravity
{\it Phys. Rev.} D {\bf 51} 10 5643--53.
 
\nonum Carlip S and Nelson J E 1994 Equivalent Quantisations of 2+1 Gravity,
{\it Phys. Lett.} B {\bf 324} 299--302. 

\bibitem{FMP} Faria Martins J and Picken R F 2007 On 2-Dimensional Holonomy {\it Preprint}  math-arXiv:0710.4310

\bibitem{gol} Goldman W M 1986 Invariant functions on Lie groups and Hamiltonian flows of surface group representations {\it Invent. Math.} {\bf 85} 263--302.

\bibitem{pick} Pick G 1899 Geometrisches zur Zahlentheorie
{\it Sitzenber. Lotos (Prague)} {\bf 19} 311--19.
\nonum Coxeter H S M 1969 {\it Introduction to Geometry}
2nd ed.(New York: Wiley) p. 209.

\bibitem{con} Connes A 1994 {\it Noncommutative Geometry} (San Diego, CA: Academic Press).

\bibitem{Turaev} Turaev V G 1991 Skein quantisation of Poisson algebras of loops on surfaces
{\it Ann. Sci. \'{E}cole Norm. Sup. (4)} {\bf 24} no. 6 635-704.


\end{thebibliography}
\end{document}